\documentclass[preprint,aps,amsmath,nofootinbib]{revtex4-1}
\usepackage{graphicx,array,dcolumn}
\usepackage{calc,tabularx, epsfig,mathrsfs}
\usepackage{hyperref}

\usepackage{amsmath,verbatim,enumerate}
\usepackage{amssymb}
\usepackage{wasysym}
\usepackage{xcolor}
\usepackage{subfigure}
\usepackage{tcolorbox}
\usepackage{multirow}
\usepackage{xspace}
\usepackage{slashed}
\usepackage{mathtools}
\usepackage{float}
\usepackage{cancel}
\allowdisplaybreaks[1]
\newlength{\figurewidth}
\newcommand{\beq}{\begin{equation}}
\newcommand{\eeq}{\end{equation}}
\newcommand{\bea}{\begin{eqnarray}}
\newcommand{\eea}{\end{eqnarray}}
\newcommand{\ba}{\begin{array}}
\newcommand{\ea}{\end{array}}

\newcommand{\ep}{\epsilon}

\newcommand{\lam}{\lambda}
\newcommand{\Lam}{\Lambda}

\newcommand{\OM}{\Omega}

\makeatother
\begin{document}
%
%%%%%%%%%%%%%%%%%%%%%%%%%%%%%%%%%%%%%%%%%%%%%%%
\title{
Note on KSW-allowability of Wine-Glass Geometry
}
\setlength{\figurewidth}{\columnwidth}
%%%%%%%%%%%%%%%%%%%%%%%%%%%%%%%%%%%%%%%%%%%%%%%
%
\author{Manishankar Ailiga}
\email{manishankara@iisc.ac.in}

\author{Gaurav Narain}
\email{gnarain@iisc.ac.in}

\affiliation{
Center for High Energy Physics, Indian Institute of Science,
C V Raman Road, Bangalore 560012, India.
\vspace{10mm}
}

%%%%%%%%%%%%%%%%%%%%%%%%%%%%%%%%%%%%%%%%%%%%%%%
\vspace{20mm}

%%%%%%%%%%%%%%%%%%%%%%%%%%%%%%%%%%%%%%%%%%%%%%%
\begin{abstract}
In this note we consider no-boundary instantons and wine-glass geometries which are of interest in the context of quantum cosmology. While the former usually appears as a dominant saddle in the path-integral, the wineglass geometry can become dominant saddle in some situations. The later has been argued to have a longer inflationary phase of the Universe. Kontsevich-Segal-Witten (KSW)-allowability criterion which classifies geometries on the basis of the requirement of having a meaningful QFT on it, pushes one to analyse the allowability of the various geometries. In this note we do a simple study to seek answer to the allowabilty of no-boundary instantons and wine-glass geometries, where the later is obtained via analytical continuation of Lorentzian deSitter in pure gravity. Our simple analysis which make use of a milder version of KSW allowability criterion shows that no-boundary instanton is KSW allowed while wine-glass geometries obtained via such analytic continuation in pure-gravity are KSW disallowed. This study however doesn't covers wineglass saddles arising in gravity coupled with matter theories. 

\end{abstract}
\vspace{5mm}

%%%%%%%%%%%%%%%%%%%%%%%%%%%%%%%%%%%%%%%%%%%%%%%

\maketitle

\tableofcontents

%%%%%%%%%%%%%%%%%%%%%%%%%%%%%%%%%%%%%%%%%%%%%%%

%%%%%%%%%%%%%%%%%%%%%%%%%%%%%%%%%%%%%%%%%%%%%%%
\section{Introduction}
\label{intro}
%%%%%%%%%%%%%%%%%%%%%%%%%%%%%%%%%%%%%%%%%%%%%%%

Understanding of gravitational physics has been significantly improved due to various studies involving Gravitational path-integrals, where they acquire center stage in investigating issues dealing with either holography or black hole physics or cosmologies. While a clean definition of path-integral in ordinary non-gravitational QFT can be arranged, such a luxury is not achievable in gravitational systems which suffers from perturbative non-renormalizability, suggesting an effective treatment of the problem \cite{tHooft:1974toh, Deser:1974nb, Deser:1974cz, Goroff:1985sz, Goroff:1985th, vandeVen:1991gw}. Besides this, the choice of contour enabling flat spacetime QFT path-integral convergent is not available. On the other hand, the conformal mode leading to unboundedness of the Euclidean gravitational path-integral is suggestive of working with the gravitational path-integral directly in the Lorentzian signature which, however, is highly oscillatory \cite{Gibbons:1978ac}. As a result, the path-integral over the geometries becomes a challenging problem which is sensitive to boundary choices and contours in the space of all possible complex metrics. 

In a semi-classical approximation, the gravitational path-integral reduces to a summation over contributions coming from various saddle geometries. Typically for given boundary choices, multiple saddle-geometries (which can be complex) appear in path-integral competing with each other. Dominance of one over the other is dictated by the external inputs, where the saddles can loose their dominance as external parameters are varied. This is analogous to phase transitions occurring in non-gravitational settings. Understanding dominance of saddles is therefore crucial in gaining a deeper insight to various phases of the gravitational systems. 

However, when the path-integral over metrics is extended to complex space, subtlety arises. Not all such metrics are \textit{allowable}, where quantum field theories can be meaningfully and sensibly defined \cite{Witten:2021nzp}. Kontsevich-Segal-Witten (KSW)-allowability criterion offers a diagnostic tool to identify such complex metrics. This criterion directly arises from the requirements of convergence and well-behavedness of the path-integral, bifurcating the space of complex geometries in to {\it allowable} and {\it non-allowable} ones. A crucial question then arises regarding the {\it allowability} of various complex saddles in the path-integral, asking piercing questions regarding the region where they lie. This however is sensitive to the choice of external parameters for example boundary configurations, in the sense that saddle allowability is affected by the boundary choices. 

In the context of quantum cosmology, it is seen that different boundary choices at initial time lead to a very different Universe in the future. By demanding a Universe to be devoid of an initial singularity, a Universe starting from zero size ({\it No-boundary Universe}) is proposed. In a simple mini-superspace model it is achievable by either fixing initial size of the Universe (Dirichlet BC) or by fixing the initial momentum (Neumann BC) \cite{DiTucci:2019bui, Narain:2021bff, Lehners:2021jmv, DiTucci:2020weq, Narain:2022msz, Ailiga:2023wzl, Ailiga:2024mmt}. In the former multiple saddles compete for dominance in various regimes given by final size of the Universe. The complex saddles which eventually becomes {\it relevant} in the path-integral computed via Picard-Lefschetz methods turn out to have unstable perturbations \cite{Feldbrugge:2017kzv, DiTucci:2018fdg,  Feldbrugge:2017fcc}. In the Neumann case, the complex saddles corresponding to Hartle-Hawking (HH) no-boundary geometries, leads to stable perturbations \cite{Lehners:2021jmv, Ailiga:2024wdx}. The saddle corresponding to the tunneling proposal is not only sub-dominant but also has unstable perturbations. Despite these appealing features, the HH-saddle however, gives rise to great puzzles: it predicts an empty Universe with least possible number of inflationary e-folds and predicts an overall spatial curvature of the Universe that is significantly larger than the observed one \cite{Maldacena:2024uhs}.

\begin{figure}[h]
    \centering
    \subfigure[\, No-boundary geometry]{ 
        \includegraphics[trim={0cm 0.8cm 0cm 0cm}, clip, scale = 0.9]{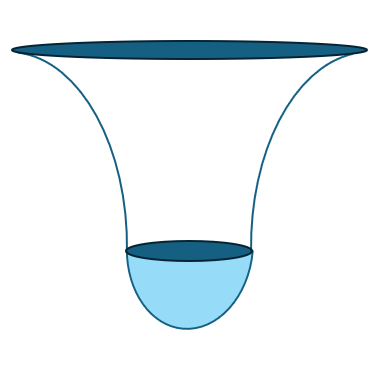}
                }
    \hspace{0.5cm}
    \subfigure[\, Wineglass geometry]{
    \includegraphics[trim={4cm 8cm 5cm 3.5cm}, clip, scale = 0.465]{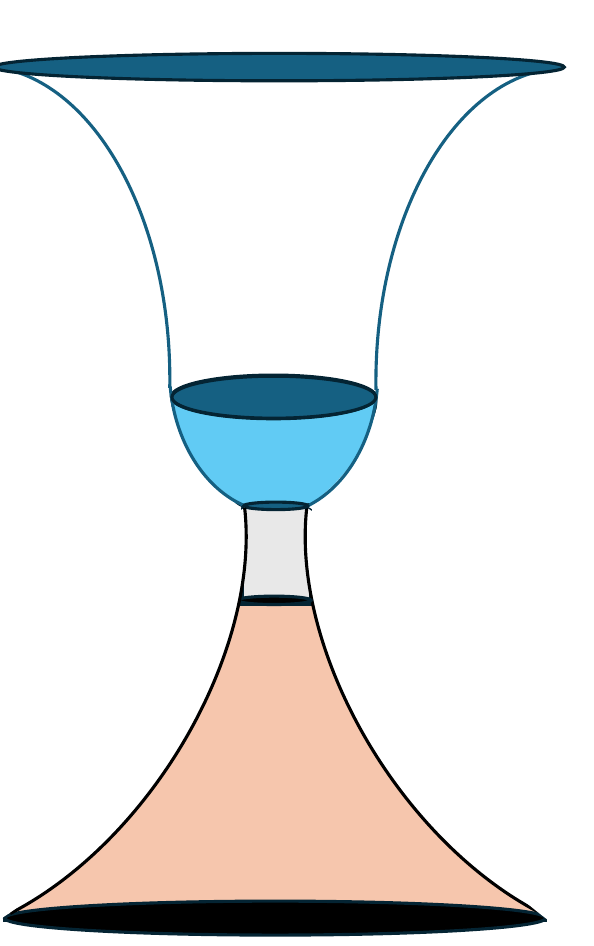}
                }
    % \caption{}
    \label{Fig:geo_plots}
\end{figure}

Recently, it has been noticed that saddle configurations corresponding to {\it wine-glass} geometries (first introduced in \cite{Lavrelashvili:1988un}), eases this tension \cite{Betzios:2024oli, Betzios:2024zhf, Betzios:2026rbv}, though such saddles may not enjoy the privilege of being dominant in the path-integral as noticed in \cite{Lavrelashvili:2026zsw}. These are wormhole saddles where the Euclidean wormhole bridges the two asymptotic regions, and are intrinsically non-perturbative configurations following from equation of motion. They are Euclidean wormhole saddles supported by Axion-field, and in situations where they are dominant saddles over others, can be analytically continued to Lorentzian signature, where they lead to cosmological evolution. A particular feature of these wormhole geometries is that they are asymptotically AdS in the far past. These gravitational saddles are noticed to overcome the various issues affecting the Hartle-Hawking no-boundary saddle i.e. supports long lasting inflationary period. Though it is still not well-established how from a single boundary configuration both wineglass saddles and no-boundary instanton can both arise, where the wineglass saddle emerges as a dominant one.

While the dust in regard to this is not settled yet, it has also been noticed recently that the two saddle configurations: Hartle-Hawking no-boundary instanton and the wineglass saddle are part of common family of Euclidean solutions, as the former can be obtained in the limit of vanishing axion charge leading to pinching of wormhole throat \cite{Lavrelashvili:2026zsw}. Presence of such wormhole configurations in the path-integral also raises sharp question regarding the factorization problem \cite{Maldacena:2004rf}. 

In this short note we check the KSW allowability of particular type of wineglass geometries that are obtained via analytical continuation of Lorentzian deSitter solution. This is important because if such geometries are dominant for some boundary configurations and are KSW-allowed then they offer a favourable proposal over the Hartle-Hawking no-boundary instanton as it overcomes various issues by giving preference to longer inflationary phase. On the other hand if they are not KSW-allowed, which may very well happen, as has been noticed for cases like bouncing cosmologies or typical wormholes geometries, then they lead to unphysical geometries on which sensible QFTs cannot be defined. It is therefore crucial to address this issue. In this short note we do a semi-rigorous study to analyse the KSW-allowability of the wineglass geometries which are analytically continued from Lorentzian dS. These, however, do not cover {\it wineglasses} which are obtained via analytic continuation of Euclidean AdS \cite{Betzios:2024oli, Betzios:2024zhf, Betzios:2026rbv, Lavrelashvili:2026zsw}.  

The rest of the paper is organized as follows: In Sec. (\ref{sec:kswc}), we briefly review the KSW criterion. Sec. (\ref{sec:NB_KSW}), details about the techniques used to check the admissibility complex geometry and particularly applied for No-boundary instanton, while Sec. (\ref{sec:WG_KSW}) deals with the admissibility of {\it wine-glass} geometry. Finally, Sec. (\ref{sec:conc}) presents summary of our findings and conclusions of the paper.

%%%%%%%%%%%%%%%%%%%%%%%%%%%%%%%%%%%%%%%%%%%%%%%
\section{KSW-criterion}
\label{sec:kswc}
%%%%%%%%%%%%%%%%%%%%%%%%%%%%%%%%%%%%%%%%%%%%%%%

The KSW criterion is proposed by Witten based on the seminal works of Louko-Sorkin (LS) and Kontsevich-Segal (KS). It is a diagonistic tool to distinguish physically admissible complex metrics from those are not \cite{Witten:2021nzp, Louko:1995jw, Kontsevich:2021dmb}. This criterion classifies a complex geometry as admissible if a well-defined quantum field theories (QFTs) formulations on the complex geometry is possible, in the sense that path-integral of the field defined on the complex background should be convergent. This requirement is simple, but in practice the KSW-criterion puts strong constraints on the metrics which are physically “allowable”. Pathological geometries such as singular bounces or wormholes with vanishing action are eliminated under this criterion \cite{Witten:2021nzp, Jonas:2022uqb}.  In the following, we briefly outline the derivation the KSW bound for diagonal complex metrics in real basis \footnote{For complex metrics that are non-diagonal to begin with, the KSW analysis in terms of eigen values is outlined in \cite{BenettiGenolini:2026raa, Krishna:2026rma}.}. 

Consider the background D($\geq 2$) spacetimes $g_{\mu\nu}$ on a manifold $\mathcal{M}$ on which we define the Euclidean path integral of a $p$-form real gauge field with field strength being a $q=p+1$-form denoted by $F$. According to the KSW-criterion, the path-integral for the $p$-form field is convergent on the given background metric if and only if the metric satisfies the following condition
\beq
\label{KSW_cri}
\begin{split}
\mathcal{I}_q[A]& =\frac{1}{2q!}\int_M d^Dx\sqrt{det\,g}\,g^{\mu_1 \nu_1}\cdots g^{\mu_q \nu_q} F_{\mu_1 \mu_2...\mu_q} F_{\nu_1 \nu_2...\nu_q} \, ,\\
g_{\mu\nu} & \, \text{is allowable iff} \,  Re\left(\sqrt{det\,g} g^{\mu_1 \nu_1}\cdots g^{\mu_q \nu_q}F_{\mu_1 \mu_2...\mu_q} F_{\nu_1 \nu_2...\nu_q}\right) > 0 \, ,
\end{split}
\eeq
for all $q \in \{0, ... , D\}$, where $\mathcal{I}_q[A]$ is the corresponding Euclidean action. For the metrics that are diagonal in the real basis, with diagonal elements $\lambda_{\mu}$ at a spacetime point $x \in \mathcal{M}$, i.e., $g_{\mu\nu} = \lambda_{\mu}(x)\delta_{\mu\nu}$, the criterion reduces to a simple pointwise condition on these diagonal components. To be more precise, for the real field strength $F$, analyzing the above constraint in the diagonal basis where the metric $g_{\mu\nu}(x)=\lambda_{\mu}(x) \delta_{\mu\nu}$, we get
\begin{equation}
\label{eq:KSW_mod_eg_conds}
    \begin{split}
        & {\rm Re}\biggl( \prod_{\mu=1}^D\sqrt{\lambda_\mu}\biggr)>0 \,\,\text{and} \\
        & {\rm Re}\biggl( \prod_{\mu=1}^D\sqrt{\lambda_\mu}\prod_{\nu\in S}\lambda_\nu^{-1}\biggr)>0\,\,\forall \, S,
    \end{split}
\end{equation}
for $q=0$ and $0<q\leq D$ respectively, where $S$ is all possible subsets of $(1, \, 2, \cdots, \,D)$. For $q=0$, the eq. (\ref{eq:KSW_mod_eg_conds}) is equivalent to
\begin{equation}
    \label{eq:q=0ksw}
    -\pi< + \arg(\lambda_1)+\arg(\lam_2)+\cdots +\arg(\lam_D)<\pi.
\end{equation}
Similarly, for other $0<q\leq D$, one finds that the above conditions are equivalent to flipping any ``$q$" number of `$+$' signs to `$-$' in front of ``$\arg$" in eq. (\ref{eq:q=0ksw}). All together, there are  $1+\sum_{q=1}^{D}{}^DC_q=2^D$ independent constraints, which are concisely expressed by the following
\beq
\label{eq:KSW_theta_conds}
-\pi < \pm a_1 \pm a_2 \pm \cdots \pm a_D  < \pi, \quad \text{where} \quad a_\mu=\arg(\lambda_\mu).
\eeq
Note that the constraints arising from a $q$-form field are identical to those from a $(D-q)$-form field, due to the identity ${}^D C_q = {}^D C_{D - q}$. Taking into account of all $0 \leq q \leq D$ forms conditions, one achieves the KSW-criterion as:
\beq
\label{eq:KSW_1}
    |a_1|+|a_2|+\cdots+|a_D|<\pi.
\eeq
or equivalently,
\beq
\label{eq:KSW_criterion}
\boxed{ \sum_{\mu=1}^D|\arg(\lambda_\mu (x) )|<\pi } \quad \forall x \in \mathcal{M}\, ,
\eeq
where $\arg (z) \in (-\pi,\pi]$ with $z$ being any complex variable. This form of KSW criterion has undergone extensive investigation, yielding several important physical implications\cite{Lehners:2021mah, Jonas:2022uqb, Hertog:2023vot, Hertog:2024nbh, Ailiga:2025fny, Lehners:2023pcn, Ailiga:2025osa}. In the subsequent sections, we investigate the KSW admissibility of the two initial condition proposal - complex geometries - utilizing the eq. (\ref{eq:KSW_criterion}).

%%%%%%%%%%%%%%%%%%%%%%%%%%%%%%%%%%%%%%%%%%%%%%%
\section{Allowability of No-boundary Instanton}
\label{sec:NB_KSW}
%%%%%%%%%%%%%%%%%%%%%%%%%%%%%%%%%%%%%%%%%%%%%%%

\begin{figure}
\centering
\subfigure[\, Hartle-Hawking instanton]{ 
\includegraphics[trim={0cm 0 0cm 0cm}, clip, scale = 0.75]{nb_geo.png}
}
\subfigure[\, HH Contour]{
\includegraphics[width=0.4\linewidth]{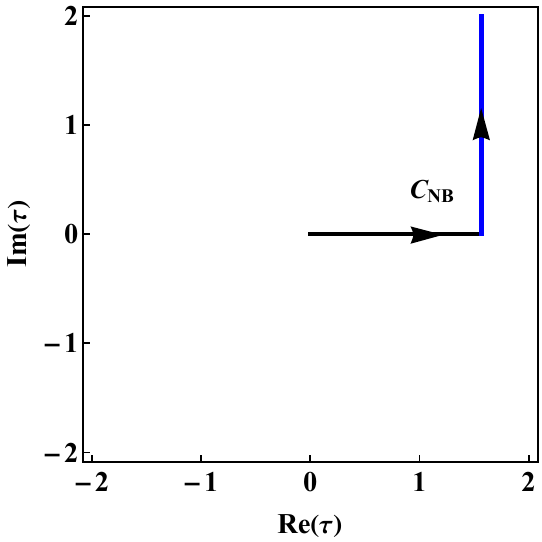}
}
\caption{A visualization of Hartle-Hawking instanton (fig (a)) and the corresponding time contour (fig (b)). In fig(a) the cyan-region denotes Euclidean sphere, while the white region represent Lorentzian deSitter. In fig(b) the horizontal line correspond to euclidean deSitter which is the hemisphere, while the vertical line correspond to Lorentzian deSitter.}
\label{Fig:HH_geo_cont}
\end{figure}

The no-boundary (NB) geometry is a half Euclidean sphere joined to the Lorentzian de Sitter space across the equator of radius $\sqrt{3/\Lam}$, see figure (\ref{Fig:HH_geo_cont}), for illustration of NB geometry. For closed Universes, the NB complex metric in Euclidean time coordinate is given as:
\beq
\label{eq:nb_geo}
ds^2 =  (3/\Lam) (d\tau^2 + \sin^2{(\tau)} d\Omega^2_{3}) \, ,
\eeq
with $\Lam >0$, being the cosmological constant. By choosing a particular contour $C_{NB}$ in complex Euclidean time plane as shown in figure (\ref{Fig:WG_plots}), one can obtain the concrete representation of NB geometry. Along the real part of $C_{NB}$ (black contour) i.e, from $Re(\tau) = 0$ to $Re(\tau)  =\pi/2$ the geometry denotes the real Euclidean geometry of a sphere. The later part of $C_{NB}$ (blue contour) i.e., from $(Re(\tau) =\pi/2, Im(\tau)=0)$ to $ (Re(\tau) =\pi/2, Im(\tau)=i y)$ , represents the Lorentzian deSitter space. This correspond to metric
\beq
\label{eq:nb_geo_ds}
ds^2 =  (3/\Lam) (-dy^2 + \cosh^2{(y)} d\Omega^2_{3}) \, ,
\eeq
which is glued at the equator $(Re(\tau) =\pi/2, Im(\tau)=0)$. The no-boundary proposal emerges as a good candidate for the ground state of the Universe, as it predicts the correct spectrum of primordial perturbations \cite{Maldacena:2024uhs, Lehners:2021jmv} and provides a natural initial condition for inflationary phase. While it still posses puzzles such as it leads to least possible number of inflationary e-folds and disagreement of overall spatial curvature of Universe with the observations. Here, we focus on studying an additional feature - KSW criterion of the no-boundary complex geometry. The admissibility of the NB geometry has been studied in earlier literature \cite{Lehners:2021mah, Hertog:2023vot, Ailiga:2025fny} and shown to be KSW allowed geometry. In the following, we briefly outline the approach used to determine the admissibility of a generic complex geometry.

For a generic time contour $\tau(u)$ that runs in the complex plane, with the real parameter $u$ with the boundary conditions $\tau(u_{0}) = \nu_{0}$ and $\tau(u_{f})= \nu_f$, the metric given in eq.(\ref{eq:nb_geo}) reads as:
\beq
\label{eq:frwmet_changed_comp_t}
{\rm d}s^2 = \frac{3}{\Lam}\left[\biggl(\frac{{\rm d}\tau}{{\rm d}u}\biggr)^2 {\rm d} u^2 
+ \sin^2{(\tau(u))} {\rm d}\OM_3^2
\right] \, .
\eeq
A generic geometry $\tau(u)$ is classified as allowable, if and only if there exists a curve $\tau(u): \nu_{0} \rightarrow \nu_{f}$ in the complex plane, such that the corresponding contour satisfies the KSW criterion along the curve, namely:
\beq
\label{eq:KSW_Ext_curv_cond}
\left|\arg\left(\tau'( u)^2\right)\right| + 3\left|\arg \sin^2{(\tau(u))}\right| < \pi \, ,
\eeq
where ($^{\prime}$) denotes derivative with respect to the $u$-parameter. For matter of convenience, let us split the KSW function into temporal and spatial parts as follows 
\beq
\label{eq:KSW_def}
\Sigma(\tau(u)) = \Sigma_{\rm temporal}(\tau(u)) +  \Sigma_{\rm spatial}(\tau(u))\, ,
\eeq
where
\begin{equation}
    \label{KSW_temp_spa_defs}
    \begin{split}
         \Sigma_{\rm temporal}(\tau(u)) = \left|\arg\left(\tau^{\prime}(u)^2\right)\right|, \;
         \Sigma_{\rm spatial}(\tau(u)) = 3\left|\arg \sin^2{(\tau(u))}\right|\, .
    \end{split}
\end{equation}
It is worth noting that the spatial KSW function depends on the position $\tau(u)$ of the time paths in the complex plane, while the temporal part depends on the tangent $\tau'(u)$ along the paths. In the following we examine the allowability of the no-boundary as well as wine-glass geometry in two steps: Ridge criterion and Extremal curve text, which we have detailed below.

%%%%%%%%%%%%%%%%%%%%%%%%%%%%%%%%%%%%%%%%%%
\subsection{Ridge Criterion}
\label{Rid_Cri}
%%%%%%%%%%%%%%%%%%%%%%%%%%%%%%%%%%%%%%%%%%

Ridge criterion technique exploits the position $\tau(u)$ dependence of time paths in spatial KSW function and offers a quick and simple verification to determine whether a complex geometry  violates the KSW bound or not. As a consequence of the absolute sign in eq. (\ref{KSW_temp_spa_defs}), one finds that the temporal and spatial components of the KSW function satisfy the inequalities $\Sigma_{\rm temporal}(\tau(u)) \geq 0$ and $\Sigma_{\rm spatial}(\tau(u)) \geq 0$, respectively. Moreover, the full KSW function obeys $\Sigma(\tau(u)) \geq \Sigma_{\rm temporal}(\tau(u))$ and $\Sigma(\tau(u)) \geq \Sigma_{\rm spatial}(\tau(u))$. Now, the above inequalities imply that the existence of a connected path from $\tau(u_{0}) = \nu_{i}$ to $\tau(u_f) = \nu_f$, along which $\Sigma_{\rm spatial}(\tau(u)) < \pi \, \forall \, u \in [u_{0},u_{f}]$ is a \textit{necessary} condition, but not \textit{sufficient} condition to satisfy the KSW bound. Therefore, if no such connected path exists — i.e., if all paths violate this necessary condition — it is a sufficient criterion to conclude that the corresponding geometry violates the KSW criterion.

%%%%%%%%%%%%%%%%%%%%%%%%%%%%%%%%%%%%%%
\subsection{Extremal Curve Test}
\label{Ext_cur}
%%%%%%%%%%%%%%%%%%%%%%%%%%%%%%%%%%%%%%

\begin{figure}
\centering
\subfigure[\, Ridge Criterion]{ 
\includegraphics[trim={0cm 0 0cm 0cm}, clip, scale = 0.7]{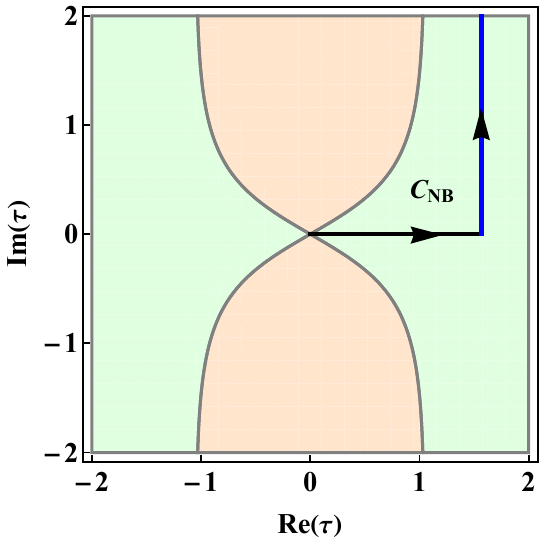}
}
\subfigure[\, Extremal Curve ]{
\includegraphics[trim={0cm 0 1.5cm 0cm}, clip,  scale = 0.58 ]{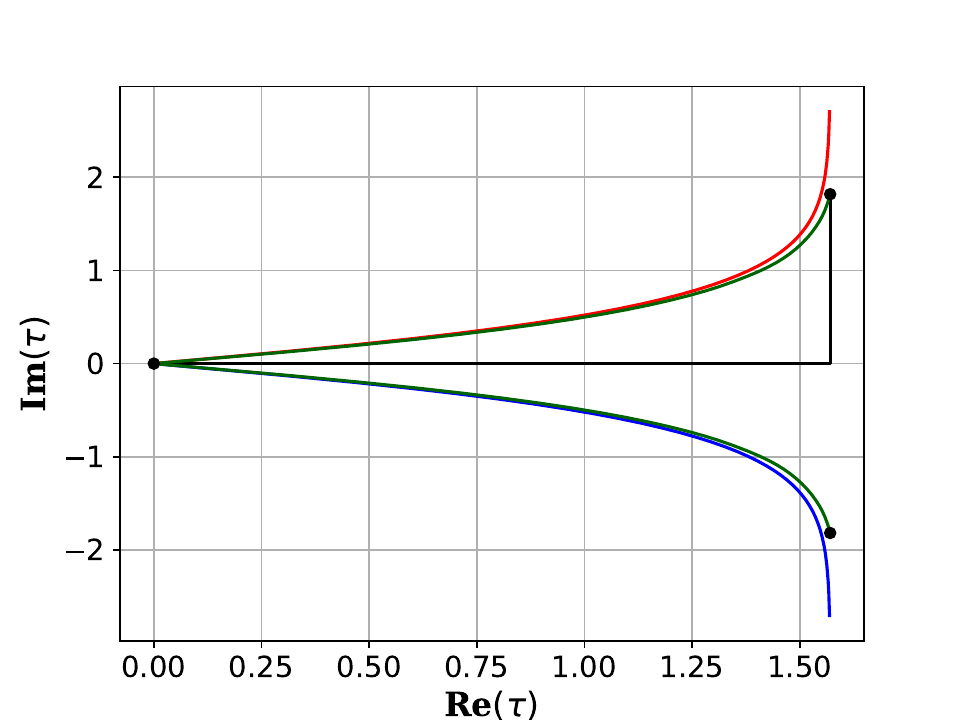}
}
\caption{KSW-allowability of Hartle-Hawking no-boundary instanton via Ridge criterion (fig a) and via a more rigorous procedure of extremal-curves (fig b). In fig a, the green region is KSW-allowed while the light orange region is KSW disallowed. In fig b, the red and blue curve denotes the extremal curves starting from $\tau=0$ and asymptote to $\mathcal{R}e(\tau) = \pi/2$. The green curve illustrates one such allowable path, along which the KSW function remains bounded by $\pi/\ep < \pi$ where $\ep \approx 1.036934$ throughout the length connecting $\nu_{0}=0$ and $\nu_f=\pi/2 \pm i 0.5$. From both fig (a) and (b) one can notice that the HH instanton is KSW allowed.}
\label{Fig:Nb_ext_all_curve}
\end{figure}

Although, Ridge criterion is robust enough for identifying KSW-disallowed points, it is not a sufficient condition to classify a geometry as admissible or not. In particular, while the absence of a path with $\Sigma_{\text{spatial}} < \pi$ is a sufficient indication of violation of the KSW bound, the presence of such a path does not guarantee the KSW bound is satisfied. Therefore, for geometries that pass the Ridge criterion, one need to perform further the conclusive check regarding their admissibility using the extremal-curve test. In contrast to the Ridge criterion, the Extremal Curve method offers a strong and conclusive classification of allowable and disallowable geometries under the KSW criterion. This check is based on the construction of an ``extremal curve" $\tau_{e}$ that saturates the inequality in eq. (\ref{eq:KSW_Ext_curv_cond}): 
\beq
\label{eq:ext_curve_1}
\left|\arg(\tau_{e}'(u))^2\right| + 3 \left|\arg \sin^2{( \tau_{e}(u))}\right| = \pi \, .
\eeq
Upon complexifying the Euclidean time as $\tau= \tau_{x} + i \tau_{y}$, the extremal equation becomes
\beq
\label{eq:ext_curve_2}
2 \left|\tan^{-1}\biggl(\frac{d\tau_{y}}{d\tau_{x}}\biggr)\right| + 3 \left|\arg \sin^2(\tau_x + i \tau_y)\right| = \pi \, ,
\eeq
which simplifies to
\beq
\label{eq:ext_curve_3}
\frac{d\tau_{y}}{d\tau_{x}}  = \pm \tan{\biggl(\frac{\pi}{2} - \frac{3}{2} \left|\arg \sin^2(\tau_x + i \tau_y)\right|\biggr)} \, ,
\eeq
with the initial conditions as $\tau_{y}(\tau_{x} = Re(\nu_0)) = Im(\nu_{0})$. One obtains extremal solutions in the complex physical time plane to the eq. (\ref{eq:ext_curve_3}), with the stated initial conditions. Then, it follows from the Petrovitsch's theorem \cite{Jonas:2022uqb, michel1901maniere}, the curves everywhere satisfying eq. (\ref{eq:KSW_Ext_curv_cond}) and starting from $\tau=\nu_0$ are constrained to remain between these two extremal curves. This suggests that an allowable curve — one that respects the KSW criterion — may exist between these extremal boundaries. Such a curve can be constructed by continuously deforming the right-hand side of eq. (\ref{eq:ext_curve_1}) and solving it. Following this, we classify a generic complex geometry as allowable, if the corresponding $\nu_f$ obeys \cite{Hertog:2023vot}: $|\tau_{y}|> |\rm Im(\nu_f)|$ at $\tau_x = \rm Re(\nu_f)$ or $|\tau_x| < |\rm Re(\nu_f)|$ at $\tau_{y}= \rm Im(\nu_f)$. The pictorial illustration of the extremal curve test is shown in the figure (\ref{Fig:Nb_ext_all_curve}).

For the no-boundary complex geometry, the initial conditions are given by $\nu_0 = 0$ and final condition is $\nu_{f} = \pi/2 + i \tau_{ds}$. We find the NB geometry passes the Ridge criterion test, see figure (\ref{Fig:Nb_ext_all_curve}). Furthermore, upon writing $\tau = \tau_x + i \tau_y$, the expression for the extremal curve is given by: \footnote{This equation leads to multiple curves, however one should consider the curve that satisfies $\tau_{y} = l \sin{\pi/8}$ when $\tau_{y} = l \cos{\pi/8}$, in limit $l \to 0,$ that are the no-boundary conditions. }
\beq
\label{eq:red_ext_curve_sol}
\cos{(3\tau_x)} \cosh{(3\tau_y)} - 9 \cos{(\tau_x)}  \cosh{(\tau_y)} + 8 = 0 \, .
\eeq
The extremal curve starts from $\tau_y(\tau_x =0) = 0$ and asymptotes the vertical line $\tau_x = \pi/2$, indicating that the curve never crosses this boundary.  As a result, the allowable condition from the Extremal curve test: $|\tau_x| < |\rm Re(\nu_f)|$ at $\tau_y = \rm Im(\nu_f)$ is achieved. Thereby ensuring the KSW allowability of the background Hartle-Hawking no-boundary saddle points. One such allowable contour for the no-boundary saddle has been shown in the Fig (\ref{Fig:Nb_ext_all_curve}).

%

%%%%%%%%%%%%%%%%%%%%%%%%%%%%%%%%%%%%%%%%%%%%%%%
\section{Allowability of Wine-Glass geometry}
\label{sec:WG_KSW}
%%%%%%%%%%%%%%%%%%%%%%%%%%%%%%%%%%%%%%%%%%%%%%%

After having studied the admissibility of the NB instanton, we now proceed to study another interesting complex geometry - Wineglass wormholes with asymptotically Euclidean AdS in the far past, see figure (\ref{Fig:geo_plots}). Although {\it wineglass} geometries can be obtained in various ways, however the one we focus on in this paper is obatined via analytic continuation of Lorentzian deSitter. These are obtained in pure gravity. It is crucial to point out that these {\it wineglass}-geometries doesn't arise directly as saddle solutions to the EH action, but rather are artificially constructed via analytic continuation of Lorentzian dS which is a saddle of Einstein-Hilbert gravity. 

Recent studies involving {\it wineglass}-geometries which are supported by matter (magnetic or axionic fields) have come across saddles which are dominant compared to others. Geometrically these saddles are Euclidean AdS which are analytically continued to lead to Lorentzian dS \cite{Jonas:2023ipa, Betzios:2024oli, Betzios:2024zhf, Betzios:2026rbv, Lavrelashvili:2026zsw}. Phenomenologically they lead to a prolonged period of inflation, potentially resolving some of the shortcomings of the no-boundary proposal. While such {\it wineglass} saddles are quite interesting, they are technically a bit involved to analyse from the perspective of {\it allowability}. 

In the present work therefore as a first attempt to analyse the KSW-allowability of {\it wineglass} geometries, we limit our attention to pure-gravity and follow the time contour approach presented in \cite{Hertog:2011ky}. In \cite{Hertog:2011ky}, utilizing the appropriate time contour, the authors showed that in the context of AdS/CFT, the no-boundary saddle admits a representation of regular Euclidean AdS domain wall that makes a smooth transition to a asymptotically deSitter. The representation discussed demonstrates the realization of an AdS-to-dS transition through a complex geometry. It is emphasized that contributions from the transiting complex geometry regulate the volume divergences in the AdS phase and account for the emergence of late-time classical behavior in the de Sitter phase. Furthermore, the regularized Euclidean AdS action of the contour gives the probability amplitude of Lorentzian deSitter history. 
\begin{figure}[h]
    \centering
    \subfigure[\, Wineglass geometry]{
    \includegraphics[trim={4cm 6.2cm 5cm 3.5cm}, clip, scale = 0.48]{Wine_glass_Geometry.pdf}
                }
    \hspace{1cm}
    \subfigure[\, Wineglass contour]{ 
        \includegraphics[trim={0cm 0 0cm 0cm}, clip, scale = 0.9]{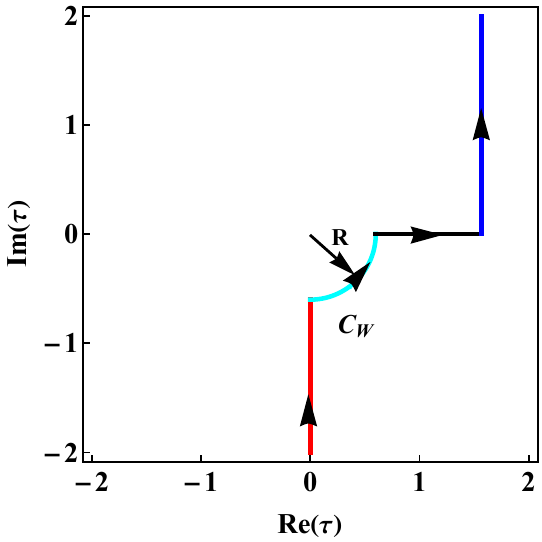}
                }
\caption{A visualization of wineglass geometry and the corresponding time contour. In fig (a): the light-orange correspond to Euclidean AdS, the gray-region corresponds to complex geometry, is the wormhole joining EAdS to EdS, the blue-region is the section of $4$-sphere which is the Euclidean deSitter, the white-region is the Lorentzian deSitter. Fig (b) correspond to the time-contour: the vertical red line correspond to EAdS, the circular region depicts the wormhole with throat, the horizontal black line correspond to section of spherical geometry, while the vertical blue line correspond to Lorenztian dS.}
    \label{Fig:WG_plots}
\end{figure}
As discussed in \cite{Hertog:2011ky}, analogous to the no-boundary case, one can realize the wineglass geometry with EAdS asymptote from metric given in eq. (\ref{eq:nb_geo}), with an appropriate choice of time contour $C_{W}$, see figure (\ref{Fig:WG_plots}). The contour $C_{W}$, divided into a part a) vertically along the negative imaginary $\tau$- axis from infinity to $\tau = -i\tau_{y}^{(0)}$ (red contour) with $\tau_{y}>0$, representing Euclidean AdS geometry, b) the circular contour from $-i\tau_{y}^{(0)}$ to $\tau_{x}^{(0)}>0$ denoting some complex geometry (cyan contour), that glues the Euclidean AdS to Euclidean sphere, c) horizontally along the real axis connecting $\tau_{x}^{(0)}$ to $\pi/2$ (black contour) depicts the Euclidean sphere that glues to Lorentzian deSitter at the equator i.e, at $\tau =\pi/2$, and d) vertically along the line to the endpoint $\tau = \pi/2+i \tau_{ds}$ (blue contour) with $\tau_{ds}>0$, representing deSitter phase (c.f. eq.\ref{eq:nb_geo_ds})). 

The geometry along part (a) is especially interesting. Along this part of $C_{W}$, i.e $\tau = -iy$ we get from eq. (\ref{eq:nb_geo}) the line element
\beq
\label{eq:EAdS_ds}
ds^2 =  (3/\Lam) (-dy^2 - \sinh^2{(y)} \, d\Omega^2_{3}) \, .
\eeq
This is a negative signature representation of the geometry of Euclidean anti-deSitter space, with a cosmological constant $\bar{\Lam} = - \Lam$. Recent studies \cite{Betzios:2024oli, Betzios:2024zhf, Betzios:2026rbv} have shown that contributions from this contour will have higher weighting in the path-integral leading to long lasting inflationary period. Now applying KSW criterion to the metric along this contour, we find
\beq
\left|\arg\left(-1\right)\right| + 3\left|\arg (-1)\right| = 4\pi > \pi \, ,
\eeq
\begin{figure}[h]
    \centering
    \subfigure[\,$q=0$ ($0$-form) KSW-criterion]{
    \includegraphics[trim={0cm 0cm 0cm 0cm}, clip, scale = 0.8]{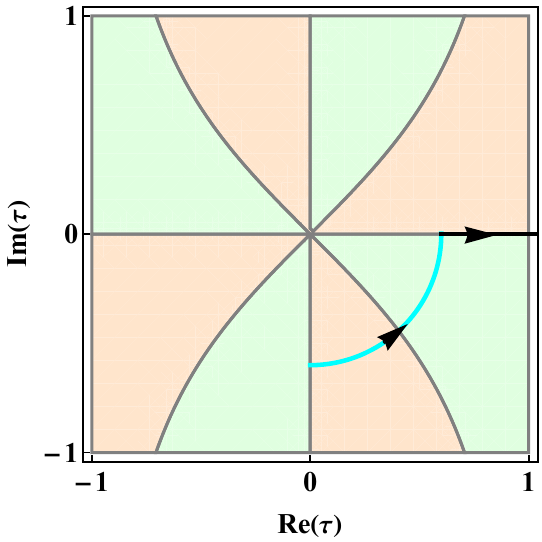}
                }
    \hspace{0.5cm}
    \subfigure[\, Ridge Criterion]{ 
        \includegraphics[trim={0cm 0 0cm 0cm}, clip, scale = 0.8]{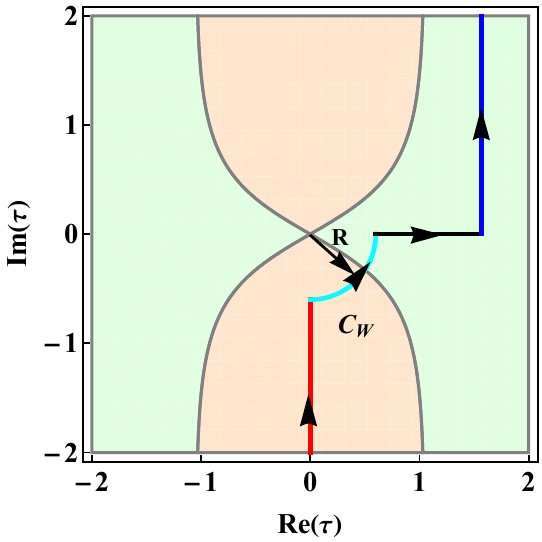}
                }
    \caption{KSW-allowability of Wine-glass instanton via $q=0$ condition (fig a) and via Ridge criterion analysis (fig b). In fig a), the green region is $\mathcal{R}e(\sqrt{g}) >0$ while light orange region denotes $\mathcal{R}e(\sqrt{g}) < 0$ region. In fig b), green region denotes Ridge criterion allowed region, while light orange represent Ridge criterion disallowed thereby KSW disallowed region. }
    \label{Fig:WG_KSW}
\end{figure}
which clearly shows the violation of KSW condition. This hints atleast in case of pure gravity the time contour realization of the wine-glass geometry is KSW disallowed, while NB saddle is KSW allowed. A better understanding can be gained by analyzing the $q=0$ admissibility condition i.e $\mathcal{R}e(\sqrt{g})>0 $, along a complex circular contour (cyan contour in fig. (\ref{Fig:WG_plots})) that glues the Euclidean AdS geometry with the Euclidean sphere. Along the circular contour, $\tau = R e^{i\theta}$, with $\theta\in [-\pi/2,0]$ the geometry given in eq. (\ref{eq:nb_geo}) reads as:
\beq
\label{eq:com_ds}
ds^2 = \frac{3}{\Lam}\left(-R^2 e^{2i\theta} d\theta^2 + \sin^2(Re^{i\theta})d\Omega_{3}^2\right) \, .
\eeq
Then the $\mathcal{R}e(\sqrt{g})$ along this contour is given by:
\beq
\label{eq:re_sq_g}
Re(\sqrt{g}) =\left( \frac{3}{\Lam}\right)^2 R \, \mathcal{R}e(i e^{i\theta} \sin^3(R e^{i\theta})) \, .
\eeq
From the plot (\ref{Fig:WG_plots}), it is clearly evident to see the violation of $q=0$ admissibility condition for some $\theta$ interval. This supports the argument that wine-glass geometry achieved via analytic continuation in pure gravity with cosmological constant is KSW disallowed. Interesting, in the vanishing throat limit i.e. $R\to0$ one achieves no-boundary instanton, which is KSW - allowed. Furthermore, for better and conclusive insight, we perform the Ridge criterion analysis. The Ridge criterion for wine-glass is illustrated in the figure (\ref{Fig:WG_plots}). The analysis reveals that the wine-glass geometry — moreover, any asymptotically EAdS configuration — violates the KSW criterion. This is because, as illustrated in Fig.(\ref{Fig:WG_plots}), the asymptotically EAdS region is surrounded by the disallowed region, implying that there exists no allowable contour connecting an asymptotically EAdS geometry to an asymptotically de Sitter configuration. This hints atleast from the time contour representation point of view, that is the wine-glass geometry obtained via analytic continuation from Lorentzian dS in pure-gravity are KSW disallowed, while NB saddle is KSW allowed.

%%%%%%%%%%%%%%%%%%%%%%%%%%%%%%%%%%%%%%%%%%%%%%%
\section{conclusions}
\label{sec:conc}
%%%%%%%%%%%%%%%%%%%%%%%%%%%%%%%%%%%%%%%%%%%%%%%

In this short paper we consider class of geometries which resemble like {\it wine-glass} and have recently gained attention in the area of quantum cosmology due to the ability of some of them to have a longer phase of inflation thereby overcoming the issues affecting the Hartle-Hawking no-boundary instanton. These wine-glasses are wormholes which typically connect the Euclidean AdS in the far past to a section of Euclidean $S^4$ which glues to Lorentzian deSitter at the equator (see figure \ref{Fig:geo_plots}). While they come under same family as the No-boundary instanton which emerges from wine-glass geometries in the limit of vanishing throat size (pinching scenario), their dominance over no-boundary instanton or vice-versa is yet to be settled \cite{Lavrelashvili:2026zsw}. 

Whether any geometry is physically meaningful in the sense whether QFTs can be defined sensibly on them, depends on whether they satisfy the KSW-allowability criterion. Geometries violating this bound are KSW disallowed and are not physically {\it relevant}. In this short paper we give a simple method of checking this bound for geometries. We apply these tools to analyse the allowability of the No-boundary instanton and find it to be KSW-allowed. While this is a known result, it gives us confidence over the usage of the tools developed to check KSW-allowability. 

We then apply these tools to the {\it wineglass}-geometries. The {\it wine-glass} geometries discussed in this paper are achieved via analytic continuation of Lorentzian deSitter solution in pure gravity. These are different from the wineglass-geometries studied in \cite{Betzios:2024oli, Betzios:2024zhf, Betzios:2026rbv, Lavrelashvili:2026zsw}, where the wormhole is supported by axionic (or magnetic) charges, and are analytic continuation of EAdS to future with Lorentzian dS. The {\it wineglass}-geomtries studied in this paper are different from that and have time contour shown in figure (\ref{Fig:WG_plots}). In our case geometries with negligible throat size correspond to contours involving sharp turns at the origin of the complex $\tau$-plane, which is singular ($R\to0$). Non-zero throat size involve contours starting at $\tau_y=-\infty$, where the geometry is EAdS. It stops at $\tau_y= - \tau^{(0)}_y <0$, leading to a non-zero throat size. The contour then ventures in complex plane to connect to No-boundary instanton contour starting at $\tau_x= - \tau^{(0)}_x>0$. The transition from EAdS to EdS happens entirely via complex geometries.  

Points lying in the far past correspond to EAdS. This mentioned in eq. (\ref{eq:EAdS_ds}) shows that all entries of the metric are negative, immediately indicating its KSW disallowability. While this is not conclusive enough, we make use of KSW allowability criterion for $q=0$-form fields. This corresponds to checking ${\rm Re}(\sqrt{g})>0$ for various points on the complex plane, thereby dividing the whole complex-$\tau$ plane into two parts. This is shown in figure (a) of \ref{Fig:WG_KSW} (green region consist of allowable geometries according to requirement ${\rm Re}(\sqrt{g})>0$, while light-orange region correspond to disallowable geometries). This is a bit crude test but gives sufficient hints. It can be clearly seen from figure (a) of \ref{Fig:WG_KSW} that the wormhole (cyan-line) lies partially in allowed and partially disallowed region. This indicates that KSW-allowability following from $0$-form field is compromised for this type of {\it wine-glasses}.

To strengthen our analysis, we check the KSW-allowability by using ridge criterion. As this test is comparatively more definitive, it offers further assurance of KSW-disallowability of class of {\it wineglass} geometries investigated in this paper. This test shows (see figure (b) in \ref{Fig:WG_KSW}) that while contours corresponding to Euclidean (black-line) and Lorentzian (blue-line) deSitter are lying in KSW allowed region, the contour (red-line) corresponding to EAdS lies in KSW-disallowed region. The contour corresponding to the wormhole (cyan-line) lies partially in allowed and partially in disallowed region. This indicates that contour corresponding to {\it wineglass} focued on in this paper are KSW-disallowed. 

One must however understand that KSW-allowability of geometries depend on quite a few things and can get affected depending on boundary choices, presence of additional fields and their interactions. The present study doesn't take those into consideration. The conclusions obtained in this paper cannot be naively generalized to all possible kind of {\it wineglass}-geometries, while it may however does provide some hint of their allowability fate which needs to be further confirmed by a proper investigation. The allowability test conducted here are for {\it wineglass} geometries in pure-gravity, which are obtained via analytic continuation of deSitter. These are seen to KSW-disallowed. A more rigorous study involving extremal curves along with presence of additional interacting fields leading to more generic types of wineglass-geometries \cite{Betzios:2024oli, Betzios:2024zhf, Betzios:2026rbv, Lavrelashvili:2026zsw} is still missing in the current work and we hope to address those in the future investigations.

%%%%%%%%%%%%%%%%%%%%%%%%%%%%%%%%%%%%%%%%%%%%%%%
\bigskip
\centerline{\bf Acknowledgements}
\vspace{3mm}
%%%%%%%%%%%%%%%%%%%%%%%%%%%%%%%%%%%%%%%%%%%%%%

We are thankful to Shubhashis Mallik for illuminating discussions on KSW and complex time contours. GN would like to thank Alok Laddha for useful discussions during the conference ``Frontiers of Gravity 2026'' held at IISER Mohali. We appreciate the ongoing ``Gauge-Gravity by Ghats'' seminar series which motivated us for this work. We are thankful to Olga Papadoulaki and Jean-Luc Lehners for clarifications. GN also acknowledges the startup support from IISc leading to workstation purchase where some of the simulations have been performed.

% %%%%%%%%%%%%%%%%%%%%%%%%%%%%%%%%%%


\begin{thebibliography}{99}

%\cite{tHooft:1974toh}
\bibitem{tHooft:1974toh}
G.~'t Hooft and M.~J.~G.~Veltman,
``One-loop divergencies in the theory of gravitation,''
Ann. Inst. H. Poincare Phys. Theor. A \textbf{20}, no.1, 69-94 (1974)
doi:10.1142/9789814539395{\_}0001
%1267 citations counted in INSPIRE as of 30 Mar 2026

%\cite{Deser:1974nb}
\bibitem{Deser:1974nb}
S.~Deser, H.~S.~Tsao and P.~van Nieuwenhuizen,
``Nonrenormalizability of Einstein Yang-Mills Interactions at the One Loop Level,''
Phys. Lett. B \textbf{50}, 491-493 (1974)
doi:10.1016/0370-2693(74)90268-8
%89 citations counted in INSPIRE as of 25 Mar 2026

%\cite{Deser:1974cz}
\bibitem{Deser:1974cz}
S.~Deser and P.~van Nieuwenhuizen,
``One Loop Divergences of Quantized Einstein-Maxwell Fields,''
Phys. Rev. D \textbf{10}, 401 (1974)
doi:10.1103/PhysRevD.10.401
%605 citations counted in INSPIRE as of 30 Mar 2026

%\cite{Goroff:1985sz}
\bibitem{Goroff:1985sz}
M.~H.~Goroff and A.~Sagnotti,
``QUANTUM GRAVITY AT TWO LOOPS,''
Phys. Lett. B \textbf{160}, 81-86 (1985)
doi:10.1016/0370-2693(85)91470-4
%516 citations counted in INSPIRE as of 30 Mar 2026

%\cite{Goroff:1985th}
\bibitem{Goroff:1985th}
M.~H.~Goroff and A.~Sagnotti,
``The Ultraviolet Behavior of Einstein Gravity,''
Nucl. Phys. B \textbf{266}, 709-736 (1986)
doi:10.1016/0550-3213(86)90193-8
%978 citations counted in INSPIRE as of 30 Mar 2026

%\cite{vandeVen:1991gw}
\bibitem{vandeVen:1991gw}
A.~E.~M.~van de Ven,
``Two loop quantum gravity,''
Nucl. Phys. B \textbf{378}, 309-366 (1992)
doi:10.1016/0550-3213(92)90011-Y
%399 citations counted in INSPIRE as of 30 Mar 2026

%\cite{Gibbons:1978ac}
\bibitem{Gibbons:1978ac}
G.~W.~Gibbons, S.~W.~Hawking and M.~J.~Perry,
``Path Integrals and the Indefiniteness of the Gravitational Action,''
Nucl. Phys. B \textbf{138}, 141-150 (1978)
doi:10.1016/0550-3213(78)90161-X
%755 citations counted in INSPIRE as of 30 Mar 2026

%\cite{Witten:2021nzp}
\bibitem{Witten:2021nzp}
E.~Witten,
``A Note On Complex Spacetime Metrics,''
[arXiv:2111.06514 [hep-th]].
%198 citations counted in INSPIRE as of 30 Mar 2026

%\cite{DiTucci:2019bui}
\bibitem{DiTucci:2019bui}
A.~Di Tucci, J.~L.~Lehners and L.~Sberna,
``No-boundary prescriptions in Lorentzian quantum cosmology,''
Phys. Rev. D \textbf{100}, no.12, 123543 (2019)
doi:10.1103/PhysRevD.100.123543
[arXiv:1911.06701 [hep-th]].
%54 citations counted in INSPIRE as of 25 Mar 2026

%\cite{Narain:2021bff}
\bibitem{Narain:2021bff}
G.~Narain,
``On Gauss-bonnet gravity and boundary conditions in Lorentzian path-integral quantization,''
JHEP \textbf{05}, 273 (2021)
doi:10.1007/JHEP05(2021)273
[arXiv:2101.04644 [gr-qc]].
%17 citations counted in INSPIRE as of 25 Mar 2026

%\cite{Lehners:2021jmv}
\bibitem{Lehners:2021jmv}
J.~L.~Lehners,
``Wave function of simple universes analytically continued from negative to positive potentials,''
Phys. Rev. D \textbf{104}, no.6, 063527 (2021)
doi:10.1103/PhysRevD.104.063527
[arXiv:2105.12075 [hep-th]].
%26 citations counted in INSPIRE as of 25 Mar 2026

%\cite{DiTucci:2020weq}
\bibitem{DiTucci:2020weq}
A.~Di Tucci, M.~P.~Heller and J.~L.~Lehners,
``Lessons for quantum cosmology from anti{\textendash}de Sitter black holes,''
Phys. Rev. D \textbf{102}, no.8, 086011 (2020)
doi:10.1103/PhysRevD.102.086011
[arXiv:2007.04872 [hep-th]].
%34 citations counted in INSPIRE as of 25 Mar 2026

%\cite{Narain:2022msz}
\bibitem{Narain:2022msz}
G.~Narain,
``Surprises in Lorentzian path-integral of Gauss-Bonnet gravity,''
JHEP \textbf{04}, 153 (2022)
doi:10.1007/JHEP04(2022)153
[arXiv:2203.05475 [gr-qc]].
%13 citations counted in INSPIRE as of 25 Mar 2026

%\cite{Ailiga:2023wzl}
\bibitem{Ailiga:2023wzl}
M.~Ailiga, S.~Mallik and G.~Narain,
``Lorentzian Robin Universe,''
JHEP \textbf{01}, 124 (2024)
doi:10.1007/JHEP01(2024)124
[arXiv:2308.01310 [gr-qc]].
%14 citations counted in INSPIRE as of 25 Mar 2026

%\cite{Ailiga:2024mmt}
\bibitem{Ailiga:2024mmt}
M.~Ailiga, S.~Mallik and G.~Narain,
``Lorentzian Robin Universe of Gauss-Bonnet Gravity,''
Gen. Rel. Grav. \textbf{57}, no.2, 29 (2025)
doi:10.1007/s10714-025-03369-2
[arXiv:2407.16692 [gr-qc]].
%9 citations counted in INSPIRE as of 25 Mar 2026

%\cite{Feldbrugge:2017kzv}
\bibitem{Feldbrugge:2017kzv}
J.~Feldbrugge, J.~L.~Lehners and N.~Turok,
``Lorentzian Quantum Cosmology,''
Phys. Rev. D \textbf{95}, no.10, 103508 (2017)
doi:10.1103/PhysRevD.95.103508
[arXiv:1703.02076 [hep-th]].
%256 citations counted in INSPIRE as of 30 Mar 2026

%\cite{DiTucci:2018fdg}
\bibitem{DiTucci:2018fdg}
A.~Di Tucci and J.~L.~Lehners,
``Unstable no-boundary fluctuations from sums over regular metrics,''
Phys. Rev. D \textbf{98}, no.10, 103506 (2018)
doi:10.1103/PhysRevD.98.103506
[arXiv:1806.07134 [gr-qc]].
%30 citations counted in INSPIRE as of 25 Mar 2026

%\cite{Feldbrugge:2017fcc}
\bibitem{Feldbrugge:2017fcc}
J.~Feldbrugge, J.~L.~Lehners and N.~Turok,
``No smooth beginning for spacetime,''
Phys. Rev. Lett. \textbf{119}, no.17, 171301 (2017)
doi:10.1103/PhysRevLett.119.171301
[arXiv:1705.00192 [hep-th]].
%166 citations counted in INSPIRE as of 30 Mar 2026

%\cite{Ailiga:2024wdx}
\bibitem{Ailiga:2024wdx}
M.~Ailiga, S.~Mallik and G.~Narain,
``Boundary choices and one-loop complex gravitational path integral,''
Phys. Rev. D \textbf{111}, no.12, 123538 (2025)
doi:10.1103/ytxr-4x1d
[arXiv:2410.19724 [gr-qc]].
%8 citations counted in INSPIRE as of 25 Mar 2026

%\cite{Maldacena:2024uhs}
\bibitem{Maldacena:2024uhs}
J.~Maldacena,
``Comments on the no boundary wavefunction and slow roll inflation,''
[arXiv:2403.10510 [hep-th]].
%61 citations counted in INSPIRE as of 25 Mar 2026

%\cite{Lavrelashvili:1988un}
\bibitem{Lavrelashvili:1988un}
G.~V.~Lavrelashvili, V.~A.~Rubakov and P.~G.~Tinyakov,
``Loss of Quantum Coherence Due to Topological Changes: A Toy Model,''
Mod. Phys. Lett. A \textbf{3}, 1231-1242 (1988)
doi:10.1142/S0217732388001483
%32 citations counted in INSPIRE as of 12 Mar 2026

%\cite{Betzios:2024oli}
\bibitem{Betzios:2024oli}
P.~Betzios and O.~Papadoulaki,
``Inflationary Cosmology from Anti-de Sitter Wormholes,''
Phys. Rev. Lett. \textbf{133}, no.2, 021501 (2024)
doi:10.1103/PhysRevLett.133.021501
[arXiv:2403.17046 [hep-th]].
%24 citations counted in INSPIRE as of 25 Mar 2026

%\cite{Betzios:2024zhf}
\bibitem{Betzios:2024zhf}
P.~Betzios, I.~D.~Gialamas and O.~Papadoulaki,
``Magnetic anti{\textendash}de Sitter wormholes as seeds for Higgs inflation,''
Phys. Rev. D \textbf{111}, no.12, 123542 (2025)
doi:10.1103/9w85-fyhs
[arXiv:2412.03639 [hep-th]].
%9 citations counted in INSPIRE as of 25 Mar 2026

%\cite{Betzios:2026rbv}
\bibitem{Betzios:2026rbv}
P.~Betzios, P.~Ghiringhelli, I.~D.~Gialamas and O.~Papadoulaki,
``A Menagerie of Wormholes and Cosmologies in the Gravitational Path Integral,''
[arXiv:2602.23432 [hep-th]].
%3 citations counted in INSPIRE as of 25 Mar 2026

%\cite{Lavrelashvili:2026zsw}
\bibitem{Lavrelashvili:2026zsw}
G.~Lavrelashvili and J.~L.~Lehners,
``Nucleating an Inflationary Universe: Euclidean Wormholes and their No-Boundary Limit,''
[arXiv:2603.11003 [hep-th]].
%2 citations counted in INSPIRE as of 25 Mar 2026

%\cite{Maldacena:2004rf}
\bibitem{Maldacena:2004rf}
J.~M.~Maldacena and L.~Maoz,
``Wormholes in AdS,''
JHEP \textbf{02}, 053 (2004)
doi:10.1088/1126-6708/2004/02/053
[arXiv:hep-th/0401024 [hep-th]].
%456 citations counted in INSPIRE as of 25 Mar 2026

%\cite{Louko:1995jw}
\bibitem{Louko:1995jw}
J.~Louko and R.~D.~Sorkin,
``Complex actions in two-dimensional topology change,''
Class. Quant. Grav. \textbf{14}, 179-204 (1997)
doi:10.1088/0264-9381/14/1/018
[arXiv:gr-qc/9511023 [gr-qc]].
%139 citations counted in INSPIRE as of 30 Mar 2026

%\cite{Kontsevich:2021dmb}
\bibitem{Kontsevich:2021dmb}
M.~Kontsevich and G.~Segal,
``Wick Rotation and the Positivity of Energy in Quantum Field Theory,''
Quart. J. Math. Oxford Ser. \textbf{72}, no.1-2, 673-699 (2021)
doi:10.1093/qmath/haab027
[arXiv:2105.10161 [hep-th]].
%187 citations counted in INSPIRE as of 30 Mar 2026

%\cite{michel1901maniere}
\bibitem{michel1901maniere}
Michel Petrovitch, M. Sur une manière d'étendre le théorème de la moyenne aux équations différentielles du premier ordre. 
{\em Mathematische Annalen}. 
\textbf{54} pp. 417-436 (1901)

%\cite{Jonas:2022uqb}
\bibitem{Jonas:2022uqb}
C.~Jonas, J.~L.~Lehners and J.~Quintin,
``Uses of complex metrics in cosmology,''
JHEP \textbf{08}, 284 (2022)
doi:10.1007/JHEP08(2022)284
[arXiv:2205.15332 [hep-th]].
%41 citations counted in INSPIRE as of 25 Mar 2026

%\cite{BenettiGenolini:2026raa}
\bibitem{BenettiGenolini:2026raa}
P.~Benetti Genolini, O.~Janssen and S.~Murthy,
``Allowable complex metrics and the gravitational index of AdS$_5$ black holes,''
[arXiv:2601.23197 [hep-th]].
%4 citations counted in INSPIRE as of 30 Mar 2026

%\cite{Krishna:2026rma}
\bibitem{Krishna:2026rma}
V.~Krishna and F.~Larsen,
``Allowable Complex Black Holes in the Euclidean Gravitational Path Integral,''
[arXiv:2602.05979 [hep-th]].
%2 citations counted in INSPIRE as of 30 Mar 2026

%\cite{Lehners:2021mah}
\bibitem{Lehners:2021mah}
J.~L.~Lehners,
``Allowable complex metrics in minisuperspace quantum cosmology,''
Phys. Rev. D \textbf{105}, no.2, 026022 (2022)
doi:10.1103/PhysRevD.105.026022
[arXiv:2111.07816 [hep-th]].
%52 citations counted in INSPIRE as of 25 Mar 2026

%\cite{Hertog:2023vot}
\bibitem{Hertog:2023vot}
T.~Hertog, O.~Janssen and J.~Karlsson,
``Kontsevich-Segal Criterion in the No-Boundary State Constrains Inflation,''
Phys. Rev. Lett. \textbf{131}, no.19, 191501 (2023)
doi:10.1103/PhysRevLett.131.191501
[arXiv:2305.15440 [hep-th]].
%27 citations counted in INSPIRE as of 25 Mar 2026

%\cite{Hertog:2024nbh}
\bibitem{Hertog:2024nbh}
T.~Hertog, O.~Janssen and J.~Karlsson,
``Kontsevich-Segal criterion in the no-boundary state constrains anisotropy,''
Phys. Rev. D \textbf{111}, no.4, 046008 (2025)
doi:10.1103/PhysRevD.111.046008
[arXiv:2408.02652 [hep-th]].
%12 citations counted in INSPIRE as of 25 Mar 2026

%\cite{Ailiga:2025fny}
\bibitem{Ailiga:2025fny}
M.~Ailiga, S.~Mallik and G.~Narain,
``Resolving degeneracies in complex {\ensuremath{\mathbb{R}}} {\texttimes} S$^{3}$ and {\ensuremath{\theta}}-KSW,''
JHEP \textbf{02}, 249 (2026)
doi:10.1007/JHEP02(2026)249
[arXiv:2507.10537 [hep-th]].
%4 citations counted in INSPIRE as of 25 Mar 2026

%\cite{Lehners:2023pcn}
\bibitem{Lehners:2023pcn}
J.~L.~Lehners and J.~Quintin,
``A small Universe,''
Phys. Lett. B \textbf{850}, 138488 (2024)
doi:10.1016/j.physletb.2024.138488
[arXiv:2309.03272 [hep-th]].
%13 citations counted in INSPIRE as of 25 Mar 2026

%\cite{Ailiga:2025osa}
\bibitem{Ailiga:2025osa}
M.~Ailiga, S.~Mallik and G.~Narain,
``Complex saddles of charged-AdS gravitational partition function,''
JHEP \textbf{02}, 054 (2026)
doi:10.1007/JHEP02(2026)054
[arXiv:2510.25396 [hep-th]].
%7 citations counted in INSPIRE as of 30 Mar 2026

%\cite{Jonas:2023ipa}
\bibitem{Jonas:2023ipa}
C.~Jonas, G.~Lavrelashvili and J.~L.~Lehners,
``Zoo of axionic wormholes,''
Phys. Rev. D \textbf{108}, no.6, 066012 (2023)
doi:10.1103/PhysRevD.108.066012
[arXiv:2306.11129 [hep-th]].
%20 citations counted in INSPIRE as of 25 Mar 2026

%\cite{Hertog:2011ky}
\bibitem{Hertog:2011ky}
T.~Hertog and J.~Hartle,
``Holographic No-Boundary Measure,''
JHEP \textbf{05}, 095 (2012)
doi:10.1007/JHEP05(2012)095
[arXiv:1111.6090 [hep-th]].
%108 citations counted in INSPIRE as of 25 Mar 2026



\end{thebibliography}
\end{document}